\documentclass[aip,cha,reprint,
 twocolumn,
 notitlepage,
 showpacs,floatfix,nofootinbib,
 superscriptaddress
 ]{revtex4-1}

 \usepackage{subfigure}
 \usepackage{amssymb}
 \usepackage{amsfonts}
 \usepackage{amsmath}
 \usepackage{amsthm}
 \usepackage{epsfig}
 \usepackage{float}
 \usepackage[usenames,dvipsnames]{color}
 \usepackage[latin1]{inputenc}

 \usepackage{graphics,graphicx,color,subfigure}
 \usepackage[hidelinks,unicode=true]{hyperref}
 \hypersetup{colorlinks=true,% Don't draw a box around links, instead color every piece of text, which is a link
 	linkcolor=blue,
 	urlcolor=blue,
 	citecolor=blue,
 	pdfhighlight=/N% When clicking a link and holding the mouse button, don't use any graphical effects - i.e. the alternative would be to behave like a "button" which is pressed within the PDF
 }%\usepackage{hyperref}
 
 \usepackage{comment}
 \usepackage[normalem]{ulem}

\newcommand{\av}[1]{\langle {#1} \rangle}

\newcommand{\add}[1]{{\color{BlueViolet} #1}}

\renewcommand{\add}[1]{{#1}}

\newcommand{\Ninf}{N_\mathrm{inf}}

\newcommand{\kmax}{k_\text{max}}
\newcommand{\kmin}{k_\text{min}}

\newcommand{\FigPathS}{./}

\begin{document}

\title{Activation thresholds in epidemic spreading with motile infectious agents on scale-free networks}

\author{Diogo H. Silva}
%\email{wesley.cota@ufv.br}
\affiliation{Departamento de F\'{\i}sica, Universidade Federal de Vi\c{c}osa, 36570-900 Vi\c{c}osa, Minas Gerais, Brazil}

\author{Silvio C. Ferreira}
%\email{silviojr@ufv.br}
%\thanks{Corresponding author.}
\affiliation{Departamento de F\'{\i}sica, Universidade Federal de Vi\c{c}osa, 36570-900 Vi\c{c}osa, Minas Gerais, Brazil}
\affiliation{National Institute of Science and Technology for Complex Systems, 22290-180, Rio de Janeiro, Brazil}

\begin{abstract} 
We investigate a fermionic susceptible-infected-susceptible model with mobility
of infected individuals on uncorrelated scale-free networks with power-law
degree distributions $P (k) \sim k^{-\gamma}$ of exponents $2<\gamma<3$. Two
diffusive processes with diffusion rate {$D$} of an infected vertex are
considered. In the \textit{standard diffusion},  one of the nearest-neighbors is
chosen with equal chance while in the \textit{biased diffusion} this choice
happens with probability proportional to the neighbor's degree. A non-monotonic
dependence of the epidemic threshold on $D$ with an optimum diffusion rate
$D_\ast$, for which the epidemic spreading is more efficient, is found for
standard diffusion while monotonic decays are observed in the biased case. The
epidemic thresholds go to zero as the network size is increased and the form
that this happens depends on the diffusion rule and degree exponent. We
analytically investigated the dynamics using quenched  and heterogeneous
mean-field theories. The former presents, in general, a better performance for
standard  and the latter for biased diffusion models, indicating different
activation mechanisms of the epidemic phases {that are rationalized} in terms of
hubs or max $k$-core subgraphs.
\end{abstract}

%\pacs{}

\maketitle

\textbf{Nowadays, we live in an interwoven world where information, goods, and
	people move through a complex structure with widely diversified types of
	interactions such as on-line friendship and airport connections.  These and many
	other systems of completely distinct nature can be equally suited in a
	theoretical representation called complex networks, in which the elements are
	represented by vertices and the interactions among them by edges connecting
	these vertices. The study of epidemic processes on complex networks represents
	one of the cornerstones in modern network science and can aid the prevention
	(or even stimulation) of  disease or misinformation spreading. The relevance of
	the interplay between diffusion and epidemic spreading in real systems is
	self-evident since hosts of infectious agents, such as people and mobile
	devices{,} are constantly moving, being the carriers that promote the quick
	transition from  a localized outbreak to a large scale epidemic scenario. In
	this work, we perform a theoretical analysis and report nontrivial roles played
	by mobility of infected agents on the efficiency of epidemic spreading running
	on the top of complex networks. We expect that our results will render impacts
	for forthcoming research related to the area.}

\section{Introduction}
\label{sec:intro}

Any system that allows an abstract mathematical representation {where} the
vertices are elements connected by edges representing interactions among  them
can be suited in the complex network framework~\cite{barabasi2016network}. A
network can be characterized by several statistical properties such as the
degree distribution probability $P(k)$ that a randomly selected vertex has $k$
contacts ($k$ is called vertex degree). Many real networks as
Internet~\cite{Pastor-Satorras2001}, actor and scientific collaboration
\cite{Ramasco2004}, airport connections~\cite{Colizza2006,Vespignani2011}
possess degree distributions with power-law tails in the
form~\cite{barabasi2016network} $P(k)\sim k^{-\gamma}$.

The importance of dynamics processes taking place on the top of a network are on an equal
footing as its structural properties~\cite{Barrat2008}. The simplest example is
a standard random walk, in which a particle lying on the vertices of the network
hops to {a} nearest-neighbor randomly selected. For connected undirected
networks, the stationary probability to find a walker on a vertex  is
proportional to its degree~\cite{Noh2004}. Since a random walk is a very basic
search mechanism, a deeper understanding of this dynamical process can aid  the
building of efficient strategies to find a specific content in a
network~\cite{Masuda2017}. Moreover, the patterns of mobility of individuals is
an issue of increasing relevance that nowadays can  be experimentally tracked
back using bluetooth and Internet~\cite{Eagle2006}, mobile
phones~\cite{Gonzalez2008}, or radio-frequency identification
devices~\cite{Cattuto2010,Stehle2011}. Another important class of dynamic
processes on networks is the epidemic spreading~\cite{Pastor-Satorras2015b}. It
is a remarkable example where an academic problem in complex systems  has turned
into applications in real processes as the forecast of outbreaks of
Ebola~\cite{Gomes2014}, H1N5 influenza~\cite{Colizza2007}, and Zika
virus~\cite{Zhang2017} to mention only a few examples. A kind of epidemic
spreading whose importance has increased significantly is the virus
dissemination {in} mobile devices~\cite{Wang2009}.

Relevance of the interplay between diffusion and epidemic spreading in real
systems is self-evident since hosts of infectious agents, such as people and
mobile devices, are constantly moving and being the carriers that promote the quick
transition from  a localized outbreak to a large scale
epidemics~\cite{Colizza2006,Colizza2007,Vespignani2011}. Diffusion has been
investigated on networks for \textit{bosonic} epidemic processes where the
vertices can be simultaneously occupied by several
individuals~\cite{Baronchelli2008} and, in particular, within the context of
heterogeneous metapopulations~\cite{Keeling2000,Colizza2007a,Colizza2007b,
	Mata2013a,GomezGardenes2018} where each vertex consists itself of a
subpopulation  and the edges represent possibility of interchange of individuals
moving from one subpopulation to another according to a mobility rule.  When
infected and healthy individuals move with the same rate on a metapopulation,
the concentration of both types is proportional to the vertex
degree~\cite{Colizza2007a}. On lattices, diffusion in bosonic models can lead to
complex outcomes such as discontinuous or continuous absorbing-state phase
transitions depending on the diffusion rates of infected and susceptible (that
can be infected) individuals~\cite{Maia2007}.

Epidemic process are commonly investigated within a \textit{fermionic} approach,
in which each vertex can host a single individual~\cite{Pastor-Satorras2015b}.
Mean-field theories predict equivalent critical properties and evolution for
bosonic and fermionic reaction-diffusion processes but they are not
identical\cite{Baronchelli2008}. One fundamental epidemic process with a
stationary active state is the susceptible-infected-susceptible (SIS)
model~\cite{Pastor-Satorras2015b} where the vertices can be in one of two
states: susceptible, which can be infected, and infected that can transmit the
infection. Infected individuals become spontaneously susceptible with rate $\mu$
while the susceptible ones in contact with $z$ infected individuals  are
infected with rate $\lambda z$. Despite its simplicity, the SIS model on
networks with power-law degree distributions presents complex behaviors and has
been subject of intensive research~\cite{Pastor-Satorras2015b,Wang2017}. Some
important features of the SIS model have been discussed in its fermionic version
as, for example, the value of the epidemic threshold above which the epidemics
lasts forever in the thermodynamical limit~\cite{Chatterjee2009}, how this limit
is approached~\cite{Ferreira2012,Castellano2010,Boguna2013,Cai2016a} and the
localization of the epidemic
activity~\cite{Mata2015,Goltsev2012,Lee2013,St-Onge2017,odor13b,Cota2016}.

The epidemic threshold of the SIS model on random graphs with power-law degree
distribution is null in the thermodynamical limit~\cite{Chatterjee2009},
irrespective of the degree exponent $\gamma$. The epidemic threshold of the SIS
model is commonly investigated using mean-field
methods~\cite{Pastor-Satorras2015b}. Two basic ones are the heterogeneous
mean-field  (HMF)~\cite{Pastor-Satorras2001} and quenched mean-field
(QMF)~\cite{Chakrabarti2008} theories. Recent reviews can be found
	elsewhere~\cite{Pastor-Satorras2015b,Wang2017}. The former considers a
compartmental approach where vertices with the same degree have the same chance
to be infected while the latter takes into account the actual structure of
network through its adjacency matrix; See section~\ref{sec:mf}. The QMF theory
is able to capture the asymptotic null threshold analytically
expected~\cite{Chatterjee2009} and observed in simulations~\cite{Ferreira2012}
for all values of $\gamma>2$ while the zero threshold happens only for
$2<\gamma<3$ in HMF.

In this paper, we investigate a diffusive fermionic SIS model where infected
agents hop to their nearest neighbors with rate $D$. Two  rules, with
\textit{(biased diffusion}) and without (\textit{standard diffusion}) tendency
to higher degree vertices are investigated. We observe that moderate diffusion
enhances epidemic activity in hubs and the threshold\footnote{Rigorously, for a
	finite system the unique asymptotic stationary state is the absorbing one where
	all vertices are susceptible~\cite{Marro2005}. {In this work},  we deal with an
	effective finite-size threshold above which the epidemic lifespan becomes
	extremely large.} asymptotically vanishes for both models while the finite-size
scaling of the threshold depends strongly on the diffusion model and the degree
exponent of the networks. For a fixed size, the standard diffusion model
presents an optimum value of $D$, in which the epidemic threshold is minimal
while a monotonic decay is found for biased diffusion. {Comparisons} between HMF
and QMF theories with the thresholds obtained in simulations on scale-free
networks with $\gamma<3$ show, in general, a higher accuracy of QMF for standard
and HMF for biased diffusion models indicating different activation
mechanisms{\cite{Castellano2012a,Cota2018}} for these mobility strategies.

The remaining of {the} paper is organized as follows. In
section~\ref{sec:models} we define the diffusive  SIS models, briefly review and
present the properties of random walks on networks {for} the investigated
mobility rules. The mean-field equations and their stability analyses are
presented in section~\ref{sec:mf}. The numerical methods are presented in
section~\ref{sec:methods}. Simulations are compared with the mean-field theories
in Sec.~\ref{sec:result}. The implications and {interpretations}  of the results
are presented in section~\ref{sec:discu}. We summarize our conclusions and
prospects in section~\ref{sec:conclu}.

\section{Models}
\label{sec:models}

The models consist of the SIS dynamics described in Sec.~\ref{sec:intro} with
rates $\lambda$ and $\mu$ on a network of $N$ vertices,  including diffusion of
infected individuals with rate $D$. The healing rate is fixed as $\mu=1$ without
loss of generality. Diffusion consists of the exchange of states between the
infected vertex and one of its nearest-neighbors selected according to a given
rule. It worths to stress that the exchange between two infected vertices does
not lead to a new state. The absence of diffusion on the susceptible vertices is
motivated  by simplification of {the} computer implementation {of the stochastic
	simulations}.

We investigated two diffusion rules. In the {standard diffusion}, the
state of an infected vertex $i$ of degree $k_{i}$ is exchanged with  a randomly
selected nearest-neighbor $j$ such that the exchange rate from vertex $i$
to $j$ is
\begin{equation}
D_{ij}=\frac{DA_{ij}}{k_{i}},
\label{eq:standard_dif1}
\end{equation}
where $D$ is the diffusion coefficient and $A_{ij}$ is the adjacency matrix
defined as $A_{ij}=1$ if $i$ and $j$ are connected and $A_{ij}=0$ otherwise. In
the {biased diffusion}, the exchange is done preferentially with
higher degree neighbors. Considering a simple linear relation $D_{ij}\propto
A_{ij} k_j$ it can be written as
\begin{equation}
D_{ij}=\frac{D A_{ij}k_{j}}{k_{i}\bar{\kappa}_{i}},
\label{eq:biased_dif1}
\end{equation}
where 
\begin{equation}
\bar{\kappa}_{i}=\frac{1}{k_i}\sum_{j}A_{ij}k_j
\label{eq:kappa}
\end{equation}
is the average degree of the nearest-neighbors of vertex $i$. This rule can
represent, for example, the mobility pattern of {people}
linked with the place where they live or work~\cite{Eagle2006}. Both models obey
the condition $\sum_jD_{ij}=D$.

The standard diffusion of a single random  walker was solved~\cite{Noh2004} and
the stationary probability that the walker is on a given vertex is proportional
to its degree. Thus, one expects that diffusion will increase the concentration
of infected individual on hubs. In the limit $D\rightarrow\infty$, the
infected walker will visit essentially the entire network implying in high
mixing where a mean-field regime is expected.

The random walk problem for biased diffusion on uncorrelated network with degree
distribution $P(k)$ can be solved using a HMF theory. Let $W_{k}$ be the
probability that the walker is at a vertex of degree $k$ that evolves as
\begin{equation}
\frac{dW_{k}}{dt}=-DW_{k}+k\sum_{k'}P(k'|k)D_{k'k}W_{k'},
\label{eq:biased_dif2}
\end{equation}
where $D_{k'k}=Dk/[k'\bar{\kappa}(k)]$ is the diffusion rate from a vertex of
degree $k'$ to a vertex of degree $k$ and $P(k'|k)$ is the probability that a
vertex of degree $k$ is connected to a vertex of degree $k'$. Assuming  absence
of degree correlations we have \cite{Boguna2003a} $P(k'|k)=k'P(k')/\av{k}$  and
$\bar{\kappa}(k)=\av{k^{2}}/\av{k}$. In the stationary state, the probability of
finding a walker on a vertex of degree $k$ is
\begin{equation}
W_{k}=\frac{k^{2}}{\av{k^{2}}}, 
\end{equation} 
where $\sum_k W_kP(k)=1$ since the walker must be somewhere on the network. This
result indicates a stronger trend  to move to the most connected vertices in
comparison with the standard diffusion.

%================================================================    
\section{Mean-field analysis} 
\label{sec:mf}

\subsection{HMF theory}	

Let $\rho_{k}$ be the density of infected vertices having  degree $k$. Dynamic
equations for this quantity are obtained including the diffusive terms in the HMF
equations of the original SIS model~\cite{Pastor-Satorras2001} and become
\begin{eqnarray}
\frac{d \rho_{k}}{dt}=&-&\rho_{k} + \lambda k(1-\rho_{k})\sum_{k'}P(k'|k)\rho_{k'}\nonumber\\
&-&{k\rho_{k}\sum_{k'}(1-\rho_{k'})D_{kk'}P(k'|k)}\nonumber\\
&+&k(1-\rho_{k})\sum_{k'}D_{k'k}P(k'|k)\rho_{k'}.
\label{eq:hmf1}
\end{eqnarray}
The first and second terms on the right-hand side  represent the healing and
infection, respectively. The third and fourth terms correspond to the diffusion
of infected vertices from or to a vertex of degree $k$, respectively, reckoning
the contribution of vertices with different degrees.

For standard diffusion, we have $D_{kk'}=D/k$. Performing a linear stability
analysis of the fixed point $\rho_{k}=0$, we obtain the Jacobian matrix
$J_{kk'}=-(1+D)\delta_{kk'} +C_{kk'}$, where
\begin{equation}
C_{kk'}=\left( \lambda +\frac{D}{k'} \right)  kP(k'|k).
\end{equation}
The epidemic threshold is obtained when the largest eigenvalue of the Jacobian Matrix is zero. 
Assuming that the network is uncorrelated, we have that $C_{kk'}$ has a positive eigenvector
$u_k=k$  with associated eigenvalue 
\begin{equation}
\Lambda_{1}=\lambda \frac{\langle k\rangle}{\langle k^{2} \rangle}+D.
\end{equation}
Since  $C_{kk'}$ is irreducible, the Perron-Frobenius theorem~\cite{newman2010networks} 
guaranties that it is the largest eigenvalue. Thus, the epidemic threshold is
\begin{equation}
\lambda_{c}=\frac{\langle k\rangle}{\langle k^{2}\rangle}.
\end{equation}
This expression  is exactly the same found for HMF theory of the non-diffusive
SIS dynamics~\cite{Pastor-Satorras2001} and does not depend  on the diffusion
coefficient $D$. The threshold vanishes for $2<\gamma<3$ and is finite if
$\gamma>3$ as {the} network size $N\rightarrow\infty$.

Considering  the biased diffusion with $D_{k k'} = Dk'/[k\bar{\kappa}(k)]$ and uncorrelated networks, the Jacobian matrix is
\begin{equation}
	J_{kk'}=-(1+D)\delta_{kk'} +\frac{D k^{2}P(k')}{\langle k^{2}\rangle} + \lambda \frac{kk'P(k')}{\langle k\rangle}.
	\label{eq:hmf3}
\end{equation}
We did not find a closed expression for the largest eigenvalue of this Jacobian
and analyzed it  using numerical diagonalization~\cite{NR}.

\subsection{QMF theory}

Let $\rho_{i}$ be the probability that the vertex $i$ is infected.
The QMF equation is also obtained introducing the diffusion terms in the
non-diffusive equation~\cite{Goltsev2012} and it becomes
\begin{eqnarray}
\frac{d\rho_{i}}{dt}=&-&\rho_{i}+ \lambda (1-\rho_{i})\sum_{j}A_{ij}\rho_{j}\\ &-&\rho_{i}\sum_{j}D_{ij}(1-\rho_{j})+(1-\rho_{i})\sum_{j} D_{ji}\rho_{j}\nonumber.
\label{eq:qmf1}
\end{eqnarray} 
The meaning of each term is analogous to those of Eq.~\eqref{eq:hmf1}.

Linear stability analysis around $\rho_{i}=0$ provides the Jacobian matrix
\begin{equation}
J_{ij}=-(1+D)\delta_{ij}+\lambda A_{ij}+{D_{ji}},
\label{eq:qmf2}
\end{equation}
where $D_{ij}$ is given by Eqs.~\eqref{eq:standard_dif1} and
\eqref{eq:biased_dif1} for standard and biased diffusion, respectively. Note
that it is a general result regardless of the correlation patterns.  The largest
eigenvalues of the Jacobians and thus the epidemic thresholds are, in general,
obtained with numerical diagonalization\cite{NR} {unless for simple graphs 
	as the one discussed in subsection~\ref{subsec:leaking}}.

\subsection{QMF theory for a leaking star graph}
\label{subsec:leaking}

Due to its importance to understand the \add{the activation of epidemic
	processes with localized activity and, in particular, the} SIS dynamics on
networks~\cite{Boguna2013,Castellano2010,Chatterjee2009}, we consider a star
graph where the center, $i=0$,  is connected to $K$ leaves, $i=1,2,...,K$. To
include the effects of the diffusion outwards the star, we assume that each leaf
has degree $k_\text{leaf}=\langle k\rangle$ to mimic a hub in a network. These
edges change the diffusion rate $D_{10}$ from a leaf to the center
($D_{j0}=D_{10}$ for $j=2,\ldots,K$) and permit that infected individuals in the
leaves leak  with rate $D_\varnothing$. Diffusion and infection from outside are
disregarded. Due to the symmetry we have that $\rho_1=\rho_2=\cdots\rho_K$ and
the $K+1$ equations are reduced to a two-dimensional system
\begin{eqnarray}
\frac{d\rho_0}{dt}= &-& \rho_0+\lambda K \rho_1(1-\rho_0) - D_{01}K\rho_0(1-\rho_1)\nonumber\\
                    &+&D_{10}K\rho_1(1-\rho_0) \label{eq:lstar1}\\
\frac{d\rho_1}{dt}= &-& \rho_1+\lambda \rho_0(1-\rho_1){-D_{10}\rho_1(1-\rho_0)}\nonumber\\
&+&{D_{01}\rho_0(1-\rho_1)}-D_{\varnothing}\rho_1 \label{eq:lstar2}.
\end{eqnarray}

A linear stability analysis around  $\rho_{i}=0$ provides the the Jacobian
\begin{equation}
\mathbb{J}=  \begin{bmatrix}
-(1+D_{01}K)     & (\lambda+D_{10})K \\ 
(\lambda+D_{01}) & -(1+D_{10}+D_{\varnothing})
\end{bmatrix}.
\label{eq:jacob_star}
\end{equation}

For standard diffusion we have $D_{01}=D/K$, $D_{10}=D/\av{k}$ and
$D_\varnothing = (\av{k}-1)D/\av{k}$ and setting the largest eigenvalue of 
the Jacobian to be zero
we obtain an epidemic threshold
\begin{equation}
\lambda_{c}=\frac{D(K+\av{k})}{2 K \av{k}} 
\left[ 
 \sqrt{1+4K\av{k}\frac{\av{k}(1+D)^{2}-D^{2}}{D^2(\av{k}+K)^{2}}}-1
 \right].
\label{eq:limiar_star_sd}
\end{equation}
Note that for $D\rightarrow 0$ and $K\gg\av{k}$, we recover the known result for QMF theory of  non-diffusive SIS on {a} star graph\cite{Ferreira2012} $\lambda_c\simeq 1/\sqrt{K}$. For $K\gg\langle k\rangle$, which represents hubs,  
and $D$ finite we obtain
\begin{equation}
\lambda_{c}\simeq  \frac{\av{k}(1+D)^2-D^2}{D K}.
\label{eq:limiar_star_qmf}
\end{equation}

The analysis is very similar for biased diffusion with the only differences that
$D_{10}=DK/[K+\av{k}(\av{k}-1))]\simeq D$ and
$D_\varnothing=\av{k}(\av{k}-1)/[K+\av{k}(\av{k}-1)]$, in which we assume that
all vertices outside the star have degree $\av{k}$. The threshold becomes
\begin{equation}
\lambda_{c}=\frac{D(K+1)}{2 K} 
\left[ 
\sqrt{1+4K\frac{2D+1}{D^2(K+1)^{2}}}-1
\right].
\label{eq:limiar_star_bias}
\end{equation}
For $K\gg \av{k}$ and $D$ finite, Eq.~\eqref{eq:limiar_star_bias} yields
\begin{equation}
\lambda_{c}\simeq  \frac{2D+1}{D K}.
\label{eq:limiar_star_qmf2}
\end{equation}

\begin{figure}[tbh]
	\centering
	\includegraphics[width=0.85\linewidth]{\FigPathS/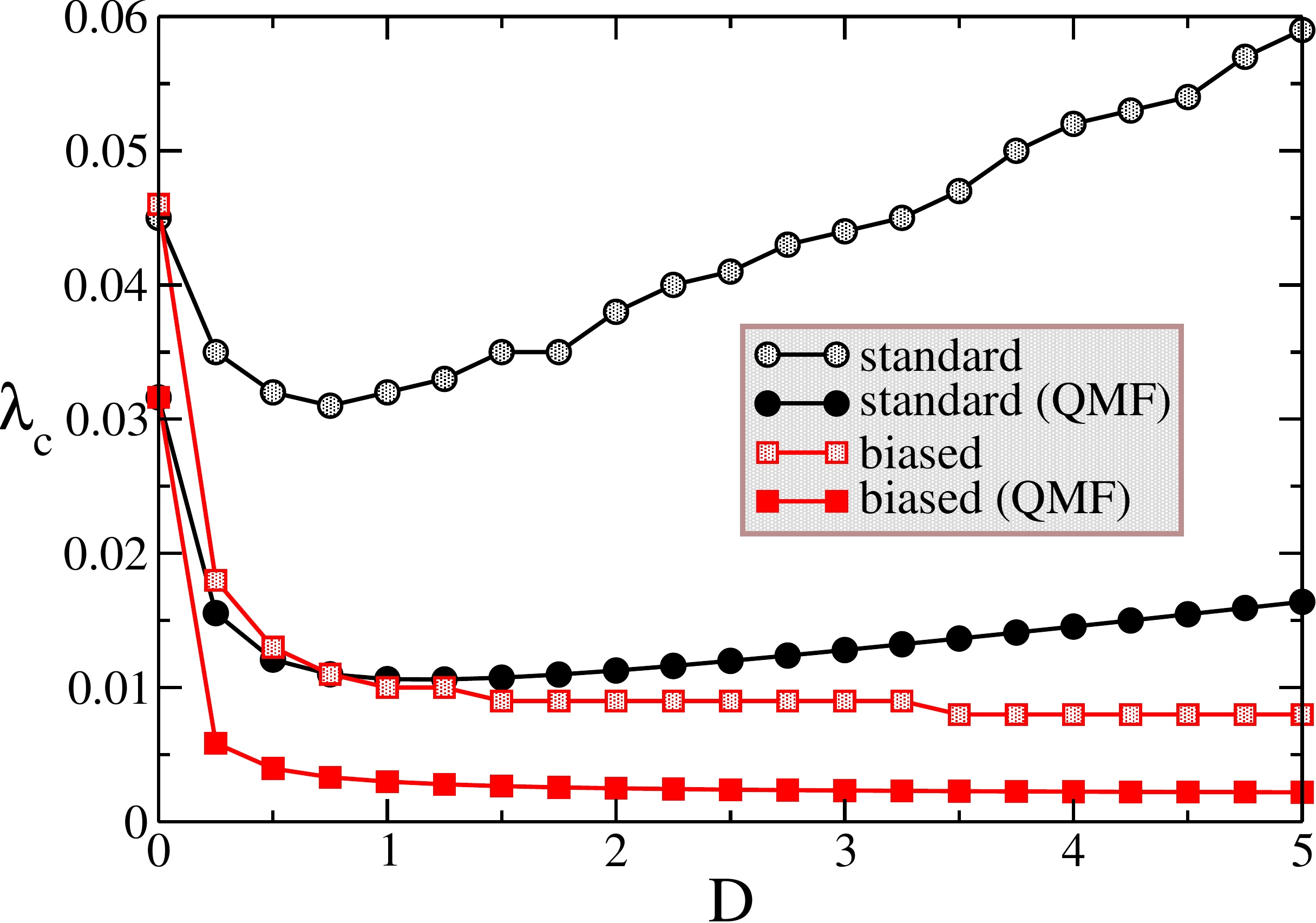}
	\caption{Epidemic threshold for SIS with standard and biased diffusion on a
		leaking star graph obtained in QMF theory,  Eqs.~\eqref{eq:limiar_star_sd} and
		\eqref{eq:limiar_star_bias}, and simulations, Sec.~\ref{sec:methods}. A center
		with $K=1000$ leaves and each leaf with $\av{k}=3$ neighbors were used.}
	\label{fig:lbcstar_random_biased}
\end{figure}

Observe that the threshold for standard diffusion presents a minimum at 
\begin{equation}
{D_\ast \simeq \sqrt{\frac{\av{k}}{\av{k}-1}}},
\end{equation}
while in biased diffusion it varies monotonically with $D$. These behaviors are 
confirmed in simulations (see Sec.\ref{sec:methods} for algorithms and
methods) on leaking star graphs shown in Fig.~\ref{fig:lbcstar_random_biased}.

\add{Taking the steady state of Eqs.~\eqref{eq:lstar1} and \eqref{eq:lstar2} we
	obtain the density of infected vertices above the epidemic
	threshold for large $K$  as 
	\begin{equation}
	\rho=\frac{\lambda-\lambda_c}{1+\lambda+\frac{\av{k}-1}{\av{k}}D}
	\end{equation} 
	and
	\begin{equation}
	\rho=\frac{\lambda-\lambda_c}{1+\lambda}
	\end{equation} 
for standard and biased diffusion, respectively. In both cases, a standard
mean-field phase transition is obtained for $K\gg 1$ where $\rho\sim
(\lambda-\lambda_c)^\beta$ with $\beta=1$ is the same exponent of the
non-diffusive case. Exploiting these expression further for $\lambda$ slightly
above $\lambda_c$, say $\lambda=a\lambda_c$ with $a\gtrsim 1$, we have that
$\rho \sim 1/K$ showing a localized  transition. These behaviors are very well
fitted by stochastic simulations (data not shown).
}

\section{Numerical methods}
\label{sec:methods}

We investigated the SIS dynamics on leaking star graphs defined in
subsection~\ref{subsec:leaking} and uncorrelated networks of size $N$ with
power-law degree distributions $P(k)\sim k^{-\gamma}$. The latter was generated
with the uncorrelated configuration model ({UCM})~\cite{Catanzaro2005}  using
lower and upper degree cutoffs $\kmin=3$  and $\kmax=\sqrt{N}$, respectively,
granting absence of the degree correlations in the scale-free regime with
$2<\gamma<3$.

The algorithm to simulate SIS model with diffusion is based on the optimized
Gillespie algorithms with phantom processes~\cite{Cota2017} that do not imply in
changes of configurations but  count for time increments. Let $N_{\text{inf}}$
be the number of infected vertices and $N_\text{e}$ the sum of the degree of all
infected vertices. In each time step, one of  the following three events is
tried.  (i) With probability
\begin{equation}
P_{\text{heal}}=\frac{\mu N_{\text{inf}}}{(\mu+D)N_{\text{inf}}+ \lambda N_\text{e}},
\end{equation}
a randomly selected vertex is spontaneously healed.
(ii) With probability 
\begin{equation}
P_{\text{inf}}=\frac{\lambda N_\text{e}}{(\mu+D)N_\text{inf}+\lambda N_\text{e}},
\end{equation}
one infected vertex is chosen with probability proportional to its degree  and
one of its nearest-neighbors is selected with equal chance.
If the neighbor is susceptible it becomes infected. (iii) Finally, with probability
\begin{equation}
P_{\text{dif}}=\frac{DN_{\text{inf}}}{(\mu +D)N_{\text{inf}}+ \lambda N_\text{e}},
\end{equation}
the states of an infected vertex and one of its nearest-neighbors  are
exchanged.  The target neighbor is chosen with equal chance in standard
diffusion and with a probability proportional to its degree  in biased
{diffusion}. The time is incremented by $dt=-\ln(u)/[(\mu+D)N_\text{inf}+\lambda
N_\text{e}]$, where $u$ is a pseudo random number uniformly distributed in the
interval $(0,1)$, while $N_{\text{inf}}$ and $N_\text{e}$ are updated
accordingly.

We apply the standard quasistationary  method~\cite{DeOliveira2005}, in which
the dynamics is reactivated to some previously visited configuration, to deal
with the absorbing state with $N_\text{inf}=0$ in a finite size system near to
the {transition} point. Implementation details can be found
elsewhere~\cite{Cota2017,Sander2016}. The quasistationary probability
$\bar{P}_n$ that the system has $n$ infected vertices near {the} transition is
computed over a time interval $t_\text{av}=10^7$ after a relaxation time
$t_\text{rlx}=10^6$. Larger or smaller values were used for very subcritical and
supercritical simulations, respectively. We  analyze the quasistationary
density $\av{\rho}$, which is the order parameter that defines active and
inactive phases, and the dynamical susceptibility~\cite{Ferreira2012}
\begin{equation} \chi= N\frac{\langle
	\rho^{2}\rangle - \langle \rho\rangle^2}{\langle \rho\rangle}.
\label{eq:sus}
\end{equation} 
whose the position of the maximum yields the effective (size-dependent) epidemic
threshold.

\section{Theory vs simulations}
\label{sec:result}

In this section we compare the thresholds obtained in the mean-field theories
with quasistationary simulations.

\subsection{Leaking star graphs}

\begin{figure}[hbt]
	\includegraphics[width=0.95\linewidth]{\FigPathS/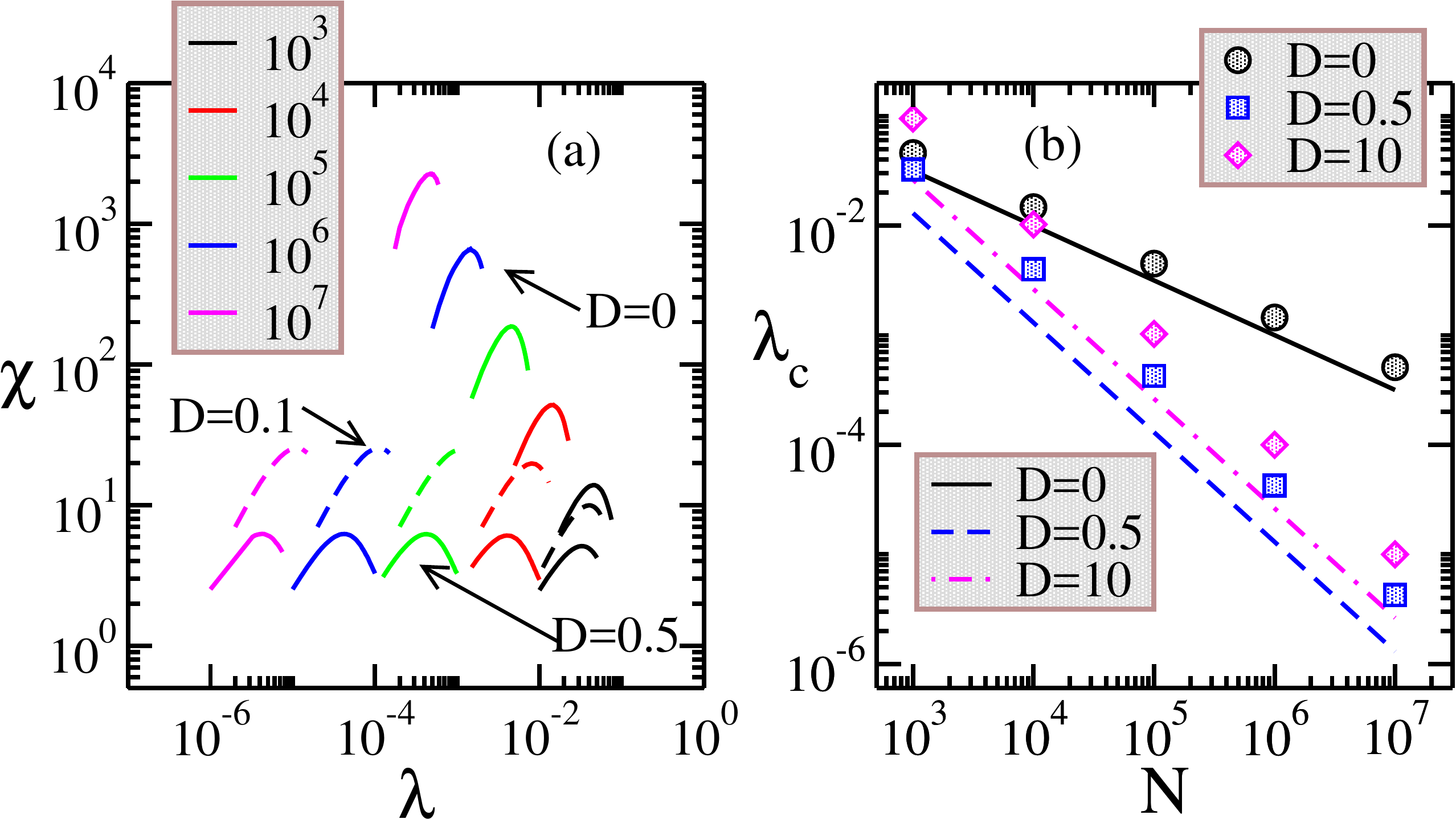}
	\caption{(a) Susceptibility as a function of the infection rate for leaking
		star graphs of different sizes (shown in the legends) with $\av{k}=3$. The
		cases without diffusion, $D=0$,  and with small {standard} diffusion rates,
		\add{$D=0.1$ (dashed lines)} and $0.5$, are shown. (b) Threshold as a function of the
		network size for different values of $D$ obtained with simulations (symbols)
		and QMF theory (lines).}
	\label{fig:star_lambdaN}
\end{figure}
The QMF predictions  given by Eqs.~\eqref{eq:limiar_star_sd} and
\eqref{eq:limiar_star_bias} are compared with simulations on leaking star graphs
in  Figs.~\ref{fig:lbcstar_random_biased} and \ref{fig:star_lambdaN}. We see
that the theory correctly predicts the qualitative dependence on $D$ for a fixed
size, Fig.~\ref{fig:lbcstar_random_biased}, as well as the scaling of the
threshold as a function of the size, Fig.~\ref{fig:star_lambdaN}(b), regardless
of $D>0$. The \add{dynamical} susceptibility \add{at the transition} diverges as the network size increases {in agreement
with a critical transition~\cite{Tauber2014}} without
diffusion~\cite{Ferreira2012,Mata2013} while it saturates in the presence of
diffusion \add{irrespective of the value of $D$ if the graph size is sufficiently large. These behaviors of the susceptibility and
	scaling of the threshold are observed in the  biased diffusion model too (data not
	shown).} \add{ The saturation happens because the center is reinfected almost
	instantaneously after it becomes susceptible since the total diffusion rate to
	the center is {proportional to} $\Ninf D\gg1 $, where $\Ninf$ is the number of
	infected leaves. The center constantly infected  depletes fluctuations in the
	number of infected individuals that determines the order parameter.}

\subsection{Power-law networks with $2<\gamma<2.5$}

\begin{figure}[h]
	\centering
	\includegraphics[width=0.84\linewidth]{\FigPathS/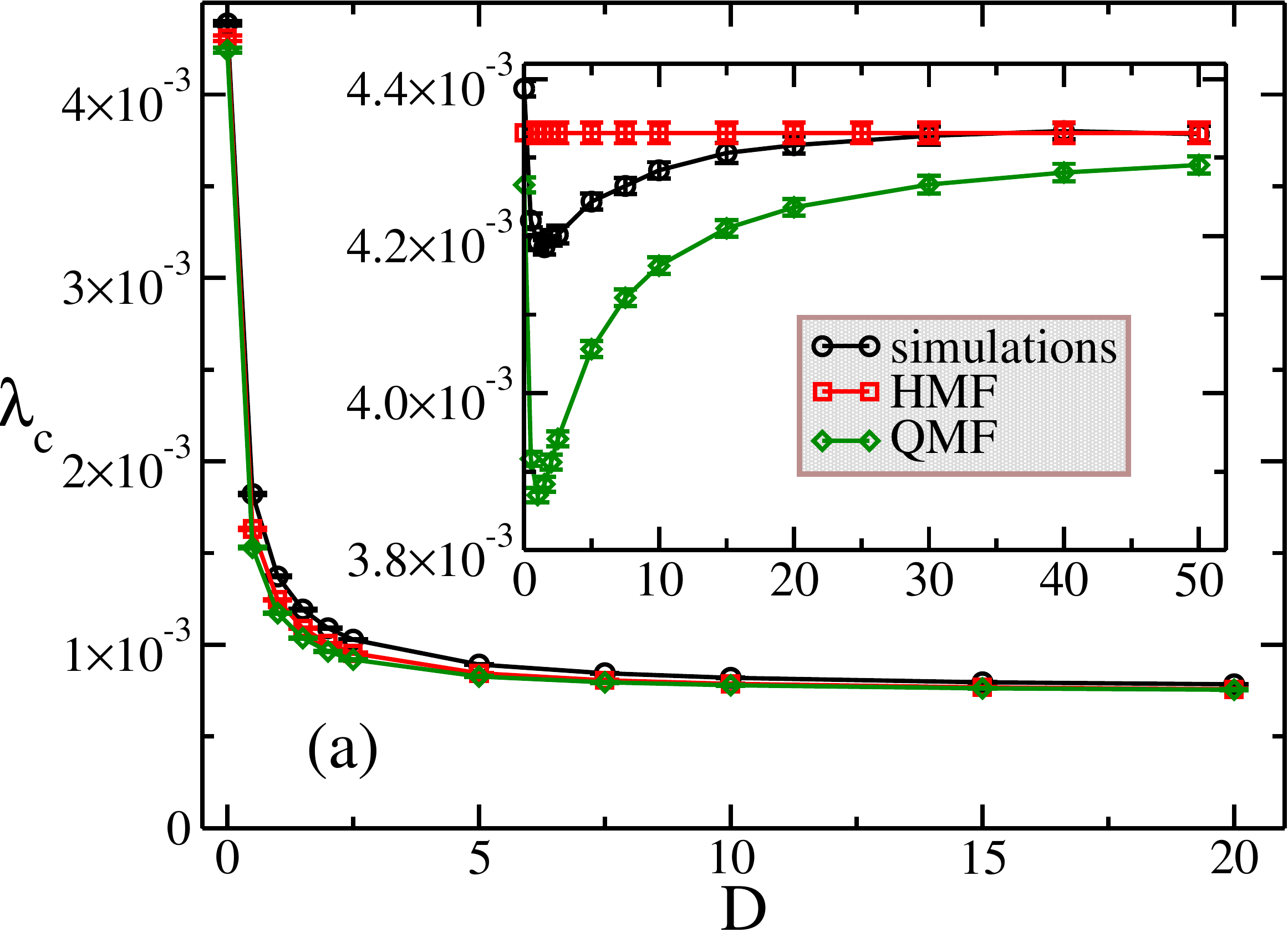}\\~\\
	\includegraphics[width=0.85\linewidth]{\FigPathS/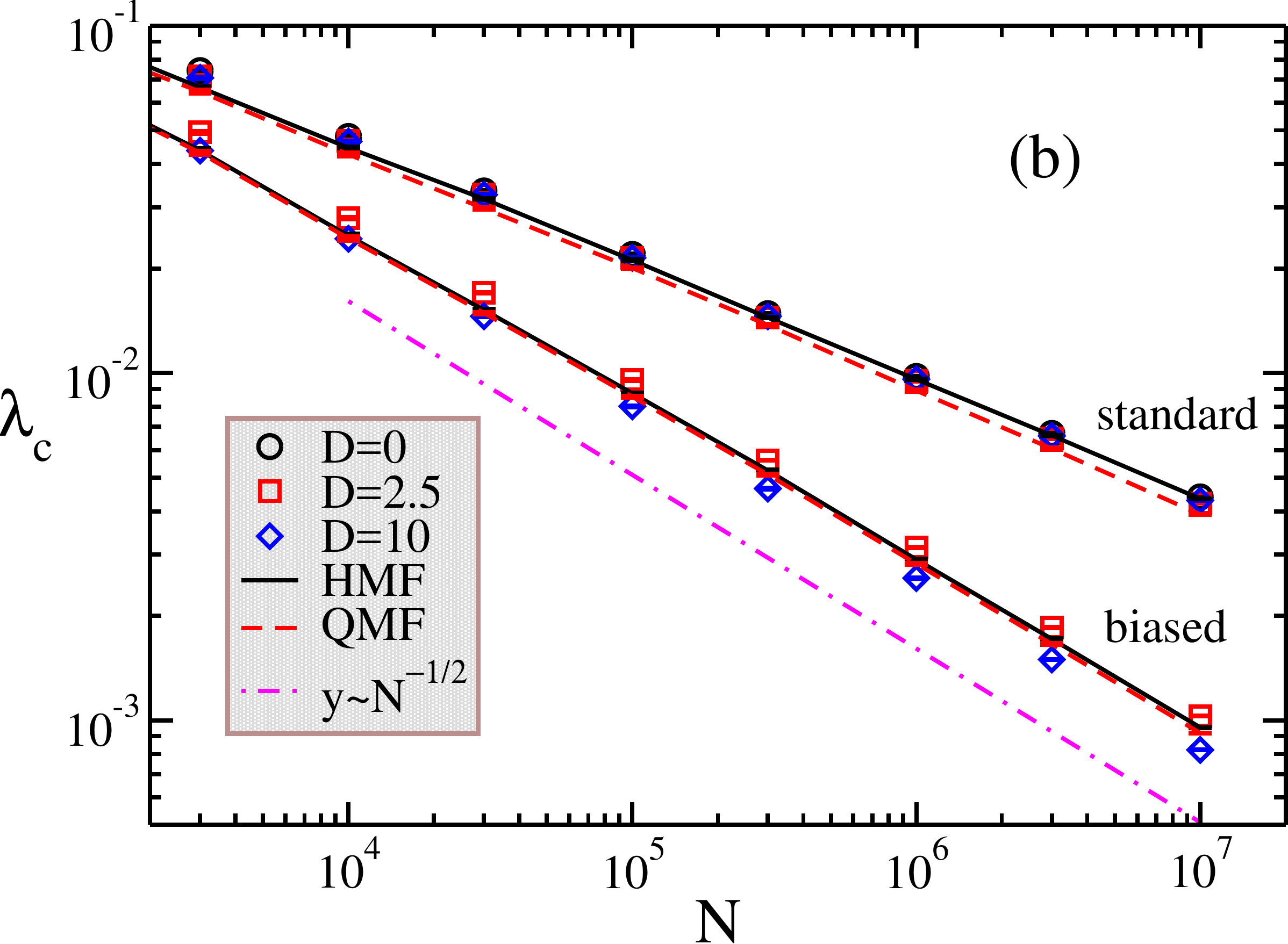}
	\caption{Epidemic thresholds for the diffusive SIS models on UCM networks with
		degree exponent $\gamma=2.25$. (a) Dependence with diffusion rate for a network
		of size $N=10^7$. Main panel shows biased while inset standard diffusion
		models. (b) Finite-size {analysis} of the epidemic {threshold} for both diffusion
		models. The scaling predicted  for a star subgraph with $K=\kmax \simeq
		\sqrt{N}$, $\lambda_c^\text{{(star)}}\sim N^{-1/2}$, is also shown.
		}
	\label{fig:lb225_thres}
\end{figure}
We compare the epidemic thresholds {of simulations} for UCM networks with
$\gamma=2.25$ with mean-field theories for both diffusion models in
Fig.~\ref{fig:lb225_thres}. The curves show a dependence with the diffusion
coefficient $D$ qualitatively described by QMF but quantitatively better fitted
by the HMF theory. Observe the narrow scale for the threshold variation in the
standard diffusion in the inset of Fig.~\ref{fig:lb225_thres}(a). We observe a
quantitative good agreement between simulations and HMF and QMF theories in both
{models}, which is evident in the finite-size analysis shown in
Fig.~\ref{fig:lb225_thres}(b). The standard diffusion presents scaling with size
in agreement with the HMF theory given by $\lambda_c^\text{(HMF)} =
\av{k}/\av{k^2}\sim \kmax^{\gamma-3}\sim N^{-0.375}$ for  $\gamma=2.25$. The
scaling for  biased diffusion is also captured by the HMF theory but it
additionally coincides with \add{QMF theory with the same scaling} of
a leaking star subgraph centered on the most connected vertex that scales as
$\lambda_c^\text{(star)}\sim \kmax^{-1}\sim N^{-1/2}$.

\subsection{Power-law networks with $2.5<\gamma<3$}

\begin{figure}[th]
\centering
\includegraphics[width=0.85\linewidth]{\FigPathS/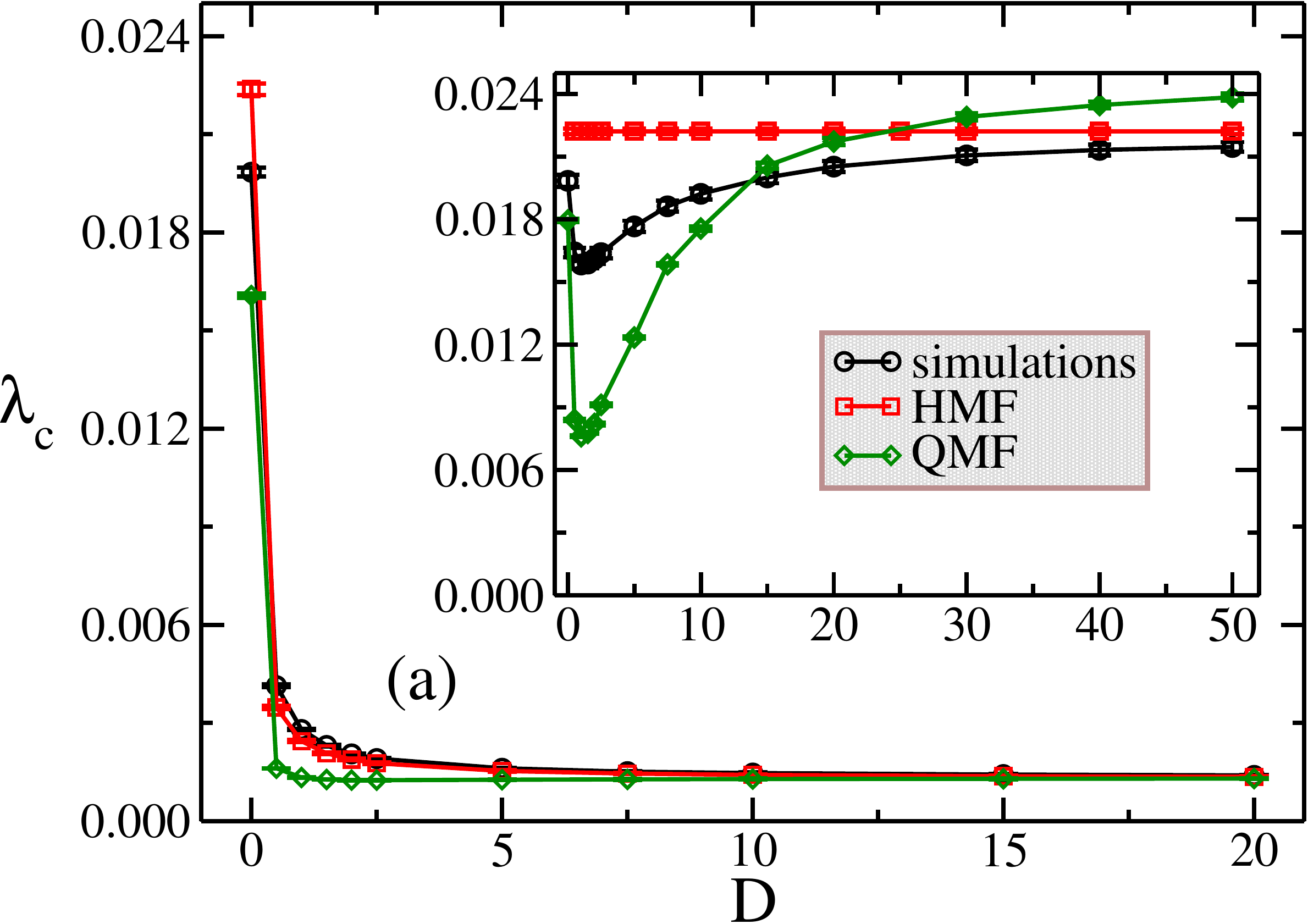}\\~\\
\includegraphics[width=0.85\linewidth]{\FigPathS/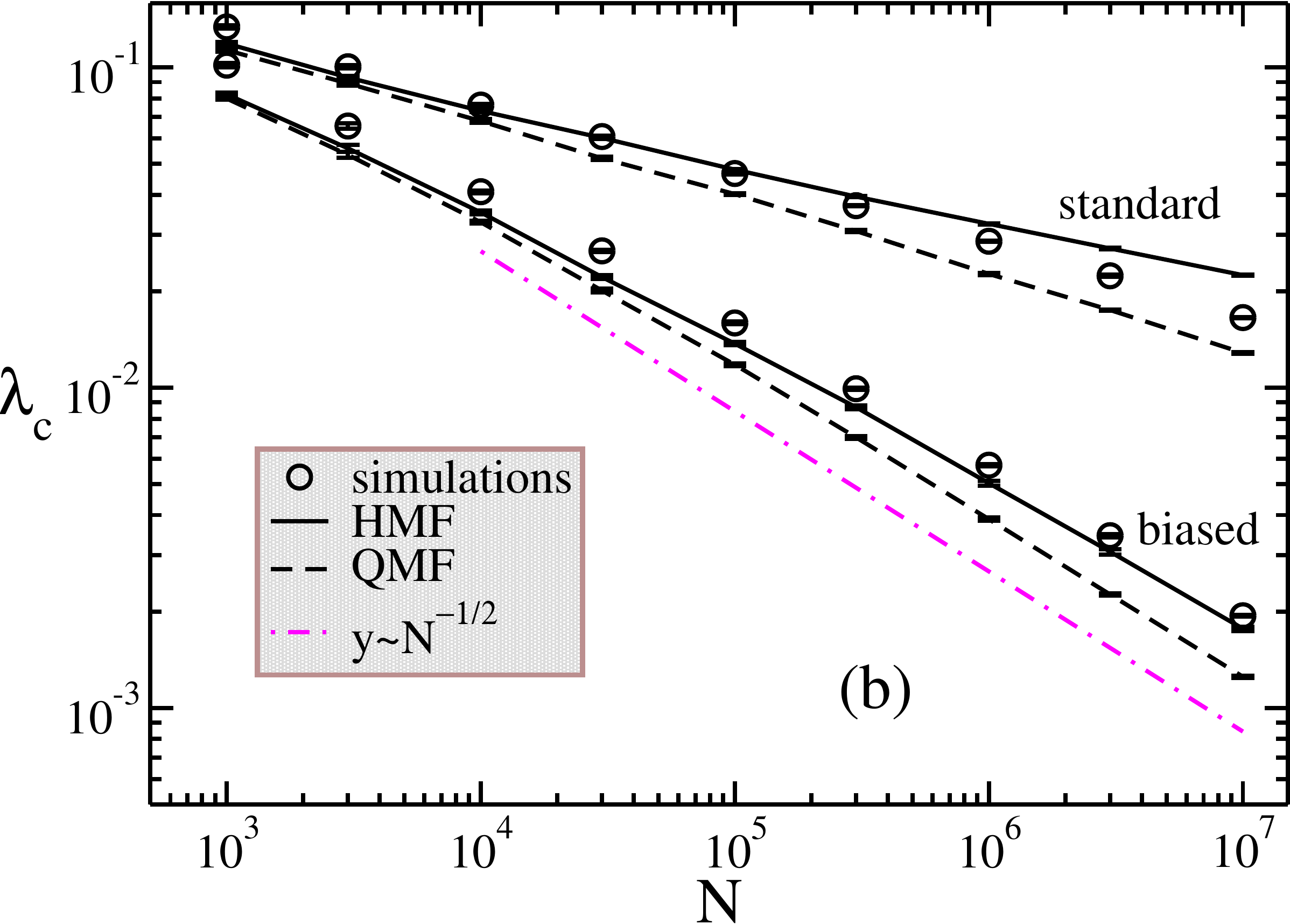}
\caption{Epidemic {thresholds} for the diffusive SIS model on UCM networks with
	degree exponent $\gamma=2.75$. (a) Dependence with the diffusion rate for a
	network of size $N=10^7$. The main panel shows  biased while the inset standard
	diffusion models. (b) Finite-size {analysis} of the epidemic {threshold} for
	both diffusion models with $D=2.5$. The scaling predicted for the epidemic
	threshold on a leaking star graph with $K=\kmax \simeq \sqrt{N}$,
	$\lambda_c^\text{(star)}\sim N^{-1/2}$, is also shown.}
\label{fig:lambdaxd_g275}
\end{figure}
The curves $\lambda_c(D)$ for UCM networks with fixed size $N=10^7$ and exponent
$\gamma=2.75$  are shown in Fig.~\ref{fig:lambdaxd_g275}(a). The qualitative
picture observed for $\gamma<2.5$ does not change with the presence of an
optimum value where the threshold is minimum in the case of standard diffusion.
Quantitatively, a considerably larger variability of the threshold as {a}
function of $D$  is found for standard diffusion. Another difference for large
$D$ is that  QMF deviates from both HMF theory and simulations. The last two
converge to each other. Theory performances  for standard and biased diffusions
can be seen in the finite-size analysis of the threshold for a fixed diffusion
coefficient in Fig.~\ref{fig:lambdaxd_g275}(b). While standard diffusion is
quantitatively better fitted by the QMF than HMF theory, as in the non-diffusive
case~\cite{Ferreira2012,Mata2013}, the biased diffusion is clearly better
described by the HMF theory, which predicts accurately both scaling and
amplitude of the epidemic threshold as a function of size. Within the
investigated size range, the threshold decay in QMF theory for the biased
diffusion  asymptotically scale as that of a leaking star subgraph centered on
the most connected vertex with degree $K\simeq\sqrt{N}$,
$\lambda_c^\text{(star)}\sim N^{-1/2}$, but this was not observed for standard
diffusion.

\section{Discussion}

\label{sec:discu}

We start our discussion {with} the epidemic threshold dependence on the diffusion
coefficient $D$ for a fixed network size. The non-monotonic dependence on $D$
for the standard and the monotonic decay for biased diffusion models regardless
of the degree exponent $\gamma$  seems to be reminiscent of the hub activation
since they qualitatively agree with the threshold obtained for leaking star
graph shown in Fig.~\ref{fig:lbcstar_random_biased} and are qualitatively
captured by the QMF theory. The role played by diffusion in the activation of
hubs is stronger in the biased case, as initially suggested by the analysis of
random walks in Sec.~\ref{sec:models}, in which a much stronger tendency to move
towards hubs is found with biased diffusion. Despite the different thresholds,
both dynamics are well described by HMF theory in the high diffusion limit,
which is expected since high mobility promotes mixing and breaks down pairwise
dynamical correlations. In such a regime HMF theory becomes exact in the
thermodynamical limit.

The finite-size scaling of the epidemic threshold in a size range $10^3<N<10^7$
shows further differences with respect to the performance of the mean-field
theories. For $\gamma<2.5$, both QMF and HMF theories have good performance for
both models. However, for $2.5<\gamma<3$ the standard diffusion is better
described by QMF while HMF presents a much better performance for biased
diffusion.

Two different mechanisms have been associated to the activation of {the}
epidemics in {the} non-diffusive SIS model on uncorrelated scale-free
networks~\cite{Castellano2012a}{. A discussion for a} more general epidemics
{can be found elsewhere~\cite{Kitsak2010}}. For $\gamma<2.5$, the epidemics in
SIS is triggered in a densely connected component of the network identified by
the maximum index of a $k$-core decomposition\footnote{A $k$-core
	decomposition\cite{Dorogovtsev2006a} consists of a pruning process that starts
	removing all vertices with degree $k_s=\kmin$ plus their edges and any other
	vertex whose degree {became} $\kmin$ after the removal, until no more vertices
	of degree $\kmin$ are present. The procedure is sequentially repeated for
	$k_s=\kmin+1,~\kmin+2$ and so on until all vertices are removed. The max
	$k$-core  corresponds to the subset of vertices and edges removed in the last
	step of the decomposition.}, hereafter called of max $k$-core. For $\gamma>2.5$
this activation is triggered in the largest hubs of the network. The epidemic
process with {max $k$-core} activation mechanism was well described by a HMF
theory, which is an intrinsically collective theory due to its  close relation
with annealed networks~\cite{Boguna2009}, while the {hub mechanism} is better
suited within a QMF theory involving a few elements of the network, namely {the}
hubs. This conjecture relating activation mechanisms and suitability of
mean-field theories has been verified for other epidemic
models~{\cite{Cota2018,Ferreira2016a}}. Applying this conjecture to the
diffusive SIS models, we have that the activation happens in the max $k$-core
for the biased diffusion in the whole range $2<\gamma<3$ and the same scheme of
the non-diffusive SIS is valid for the standard diffusion, with hub triggering
activation for $\gamma>2.5$ and max $k$-core for $\gamma<2.5$. \add{Table~\ref{tab:scheme} summarizes the distinct
	frameworks observed in our analysis of the diffusive SIS models with respect to
	the most suitable mean-field theory and the triggering activation mechanisms
	in each case.}

\begin{table}[ht]
	\centering
	\def\arraystretch{1.5}%\setlength{\tabcolsep}{0.3em}
	\add{
	\caption{Schematic summary of the basic properties of diffusive SIS models for 
		two regimes of scale-free networks.}
	\label{tab:scheme}
	\begin{tabular*}{\linewidth}{|c @{\extracolsep{\fill}}|c|c|c|c|}
		\toprule
		Model     & & {$2<\gamma<2.25$}       & & {$2.5<\gamma<3$}\\ \cline{1-1}\cline {3-3}\cline{5-5}
		& & HMF and QMF work       & & QMF outperforms HMF\\
		standard~ & & max $k$-core activation & & hub activation  \\\cline{1-1} 
		\cline {3-3}\cline{5-5}
		%& &  &   &           \\         
		& & HMF and QMF work       & & HMF outperforms QMF \\
		biased    & & max $k$-core activation & & max $k$-core activation \\
		\toprule
	\end{tabular*}
}
\end{table}

We could naturally wonder why do biased and standard diffusion {models} behave
so differently if  mobility favors localization in hubs for both cases? A high
diffusivity implies that the infected individual stays shortly in a same vertex.
In standard diffusion, when the infected individual {leaves} a hub {towards} a
{randomly selected} neighbor, it more probably arrives at a low degree vertex,
present in a much larger number, where it spreads {the} infection less
efficiently. So, even enhancing mobility to higher degree vertices, the net
effect of {sufficiently high} standard diffusion is to reduce the infection
power of hubs.  In the biased diffusion, the mobility towards hubs is highly
favored such that the infected individual stays moving mostly among hubs
keeping{, therefore,} its spreading efficiency at high levels. Moreover, these
high degree vertices belong to the max $k$-core in random scale-free networks in
consonance with the activation mechanism for $2.5<\gamma<3$ in the biased
diffusion model.

\section{Concluding remarks}
\label{sec:conclu}

Mobility is a fundamental promoter of epidemic spreading in real world and its
effects on epidemic models on networks have been dealt in the context of bosonic
processes where a single vertex can host several infected
individuals~\cite{Colizza2007a}. A lot of theoretical attention has been
dedicated to the  investigation of fermionic epidemic models where no more than
one individual can lay on a vertex~\cite{Pastor-Satorras2015b}. However, the
effects of mobility in the fermionic epidemic models on networks have been not
addressed thoroughly. We consider the role played by mobility in the phase
transition of two fermionic diffusive SIS models on scale-free networks with
degree distribution $P(k)\sim k^{-\gamma}$ and exponent $2<\gamma<3$. Both
standard and biased diffusion models were considered. In the latter{,} higher
degree vertices are favored.

The epidemic thresholds were investigated on large networks using stochastic
simulations and compared with QMF and HMF theories. Biases reduce significantly
the epidemic threshold for a fixed network size, and is very well described by
the HMF theory. Standard diffusion {yields} a non-monotonic dependence on $D$
with an optimum value {$D_*$} where the threshold at a given size is minimum.
QMF theory describes better the epidemic threshold for standard diffusion.
Different triggering mechanisms of the epidemic phase were identified. While
biased diffusion leads to a max $k$-core activation for all values of
$2<\gamma<3$, the standard diffusion behaves as the non-diffusive case, being
activated by hubs for $\gamma>2.5$ and max $k$-core for $\gamma<2.5$.

\add{Outstanding examples of theoretical studies  for epidemic process on
	networks~\cite{Pastor-Satorras2001,Pastor-Satorras2015b} currently guide applied
	research in predicting real-world epidemic outbreaks. Therefore, understanding
	better the role of mobility in epidemic processes can aid the improvement and
	accuracy of these realistic models.} Our study reveals non-trivial effects of
{the} mobility on the outcomes of epidemic models running  on  the top of
heterogeneous networks. We expect that it will render impacts for forthcoming
research in the field.  As prospects, the case $\gamma>3$ needs further
attention. Our first analysis points out a very strong depletion the
fluctuations, markedly changing  the nature of the epidemic transition. Still,
from the {application} point of view, the role of correlation and assortativity
patterns, which are ubiquitous in real networked systems,   as well as the
mobility of susceptible individuals also needs additional investigations.
\add{Finally,  the inclusion of dynamical correlations using heterogeneous
	pairwise approximations~\cite{Mata2013,Mata2014} can provide more accurate
	predictions in forthcoming studies.}

\begin{acknowledgments}
	%We thank discussions with G. \'Odor and J. A.  Hoyos.
	This work was partially supported by the Brazilian agencies CAPES, CNPq and
	FAPEMIG. S.C.F. thanks the support from the program  \textit{Ci\^encia sem
		Fronteiras} - CAPES under project No. 88881.030375/2013-01.
\end{acknowledgments}

%\bibliography{SISdiff}

%merlin.mbs aipnum4-1.bst 2010-07-25 4.21a (PWD, AO, DPC) hacked
%Control: key (0)
%Control: author (8) initials jnrlst
%Control: editor formatted (1) identically to author
%Control: production of article title (0) allowed
%Control: page (1) range
%Control: year (1) truncated
%Control: production of eprint (0) enabled
%

\end{document}